\documentclass[pdflatex,sn-vancouver-ay]{sn-jnl}

\usepackage{graphicx}
\usepackage{multirow}
\usepackage{xcolor}
\usepackage{textcomp}
\usepackage{amsmath}
\usepackage{longtable}
\usepackage{geometry}
\usepackage{array} 

\title{\textbf{A moving target in AI-assisted decision-making: Dataset shift, model updating, and the problem of update opacity}}

\author*{\fnm{Joshua} \sur{Hatherley}}\email{jjh@hum.ku.dk}

\affil{\orgdiv{Center for the Philosophy of AI}, \orgname{University of Copenhagen, Denmark}}

\abstract{Machine learning (ML) systems are vulnerable to performance decline over time due to dataset shift. To address this problem, experts often suggest that ML systems should be regularly updated to ensure ongoing performance stability. Some scholarly literature has begun to address the epistemic and ethical challenges associated with different updating methodologies. Thus far, however, little attention has been paid to the impact of model updating on the ML-assisted decision-making process itself, particularly in the AI ethics and AI epistemology literatures. This article aims to address this gap in the literature. It argues that model updating introduces a new sub-type of opacity into ML-assisted decision-making – update opacity – that occurs when users cannot understand how or why an update has changed the reasoning or behaviour of an ML system. This type of opacity presents a variety of distinctive epistemic and safety concerns that available solutions to the black box problem in ML are largely ill-equipped to address. A variety of alternative strategies may be developed or pursued to address the problem of update opacity more directly, including bi-factual explanations, dynamic model reporting, and update compatibility. However, each of these strategies presents its own risks or carries significant limitations. Further research will be needed to address the epistemic and safety concerns associated with model updating and update opacity going forward.

\bigskip

This is a pre-print of: Hatherley, Joshua. 2025. A moving target in AI-assisted decision-making: Dataset shift, model updating, and the problem of update opacity. \textit{Ethics and Information Technology} 27(2): 20. \href{https://doi.org/10.1007/s10676-025-09829-2}{10.1007/s10676-025-09829-2}}

\begin{document}

\maketitle

\section{Introduction}

Machine learning (ML) systems are increasingly being implemented to assisted human decision-makers in high-stakes environments including healthcare, finance, and warfare \citep{mckernan2024machine,singh2022reinforcement,sparrow2019promise}. A critical weakness of ML systems, however, is that they are vulnerable to performance decline over time due to dataset shift, which occurs due to mismatches between the data on which a model was trained and the data on which it is used \citep{quinonero2022dataset}. In response, researchers often suggest that ML systems should be regularly updated to ensure stability or improvement in the performance of these systems with the passing of time \citep{finlayson2021clinician,guo2021systematic}, and some scholarly literature has begun to appear that addresses emerging challenges and concerns associated with different model updating methodologies \citep{adam2022error,feng2022clinical,jenkins2021continual,pruski2023ethics}. Thus far, however, little attention has been directed toward investigating the potential impact of model updating on the ML-assisted decision-making process itself, particularly in the ethics and philosophical literature concerning AI.

This article aims to address this gap in the literature. It argues that model updating introduces a new sub-type of opacity into ML-assisted decision-making – update opacity – that occurs when users cannot understand how or why an update has changed the reasoning or behaviour of an ML system. This type of opacity presents a variety of distinctive epistemic and safety concerns that available solutions to the black box problem in ML are largely ill-equipped to address. For example, update opacity is likely to exacerbate performance drops in human-ML teams that result from model updating, and to increase the complexity of reconciling human judgments with algorithmically generated outputs. A variety of alternative strategies may be developed or pursued to address the problem of update opacity more directly, including bi-factual explanations, dynamic model reporting, and update compatibility. However, each of these strategies has significant limitations or presents its own risks. Further research will be needed to address the epistemic and safety concerns associated with model updating and update opacity going forward.

The article is structured as follows. Section \ref{2} gives an overview of model updating, model updating frameworks, and some of the epistemic and safety concerns associated with model updating. Section \ref{3} provides an overview of existing literature concerning opacity and the black box problem in ML. Section \ref{4} introduces the concept of update opacity and describes the problem of update opacity in ML-assisted decision-making. Section \ref{5} analyses strategies for addressing the problem of update opacity going forward. Section \ref{6} summarises the findings of the article and suggests some avenues for further research. 

\section{The model updating imperative}\label{2}

In this section, I discuss the promise and perils of model updating in high-stakes, ML-assisted decision-making. Section \ref{2.1} provides an overview of the causes and consequences of dataset shift. Section \ref{2.2} discusses three different approaches for updating ML systems: conventional, dynamic, and continuous. Section \ref{2.3} outlines some key epistemic and safety concerns associated with model updating. Section \ref{2.4} summarises and concludes section \ref{2}. 

\subsection{Dataset shift}\label{2.1}

Dataset shift “occurs when a machine-learning system underperforms because of a mismatch between the data set with which it was developed and the data on which it is deployed” \citep[283]{finlayson2021clinician}.\footnote{The overview of dataset shift in this subsection is heavily indebted to \cite{finlayson2021clinician}} Technological changes, for example, can disrupt the consistency of data collection methods, which, in turn, can affect the performance of an ML model. Consider a diagnostic ML system that relies on data from patients' electronic health records (EHRs). If the software used to collect or organise EHR data is updated or replaced (e.g., a shift to a new version of the software or a complete overhaul of the system), the new data collected may be structured or formatted differently from the training data, even if the underlying information remains the same. This can introduce inconsistency, leading the model to misinterpret or struggle to handle the new inputs, causing a drop in performance.

Dataset shift can also occur due to population or environmental changes, particularly in domains like healthcare, where the characteristics of the population served by a system can evolve over time. This form of dataset shift occurs when the distribution of inputs changes because of demographic, social, or environmental factors. In healthcare, for instance, if a hospital's patient demographic shifts due to gentrification (e.g., a wealthy population moving into a historically lower-income area), the model may encounter different disease patterns, medical conditions, and health profiles that were not represented in the original training data. This mismatch between the trained data and the new patient population can impair the model’s ability to provide accurate predictions or diagnoses.

Finally, behavioural changes can also drive dataset shift. Suppose, for example, a celebrity is diagnosed with a rare condition, leading to increased media coverage and heightened public awareness. As a result, more patients seek medical attention for this condition, altering patient demographics and the frequency with which this condition is diagnosed. A diagnostic model that was trained on data from a time when the condition was less common may now be less effective at handling the sudden influx of patients, since this shift in patient behaviour was not reflected in the model’s training data.

\subsection{Model updating}\label{2.2}

One of the most commonly suggested ways to address dataset shift is through model updating \citep{finlayson2021clinician,guo2021systematic}. Model updating involves fine-tuning or retraining an existing model on new data to address performance issues, to maintain performance stability, or indeed, to improve model performance. While the debate about how best to go about such updating is ongoing \citep{adam2022error}, at least three different updating frameworks have been proposed in the literature: conventional, dynamic, and continuous \citep{jenkins2021continual} (see Table \ref{tab:tab1}). 

\begin{longtable}{|>{\raggedright}p{3.8cm}|>{\raggedright}p{2.4cm}|>{\raggedright}p{1.8cm}|>{\raggedright}p{2cm}|}
    \hline & \textbf{Conventional} & \textbf{Dynamic} & \textbf{Continuous}\\
    \hline Update frequency & Infrequent & Regular & Regular \\
    \hline \textit{Performance monitoring} & Ad hoc & Continuous & Continuous \\
    \hline \textit{Data collection} & Ad hoc & Ad hoc & Ongoing\\
    \hline
    \caption{Overview of conventional, dynamic, and continuous updating frameworks for ML systems.}
    \label{tab:tab1}
\end{longtable}

Conventional updating involves ad hoc updates to address specific performance issues identified in an ML system. Performance monitoring or surveillance are minimal. Potential performance issues may be flagged by the users of a system or identified during the occasional technical audit. Training data for updates is also collected on an ad hoc basis, and each update is performed as a discrete operation either at regular intervals or only when opportunity allows. Dynamic updating differs from conventional updating insofar as performance monitoring and surveillance are conducted on an ongoing basis. ML systems are subject to monitoring strategies such as real-time performance monitoring, drift and anomaly detection, and bias auditing. However, updates are still performed as discrete operations to address specific issues, either at regular intervals or only when the opportunity allows, and data collection is performed on an ad hoc basis. Finally, continuous updating differs from dynamic updating insofar as it involves establishing a ongoing feedback loop between performance surveillance and model updating that are each conducted on a continuous and ongoing basis. New data is fed back into the model, not only for ongoing performance maintenance, but also for ongoing performance improvement. Data collection is proactive and incorporated into standard practice to provide these systems with a consistent and ongoing flow of new training data. Continuous updating may also blur the boundary between research and practice and warrant stronger regulatory oversight in some industries, such as healthcare \cite[see][]{sparrow2024should}.

\subsection{Implications}\label{2.3}

As discussed, model updating can mitigate dataset shift and enable ongoing performance improvement in ML systems. However, model updating also presents a variety of epistemic concerns and challenges for human-AI interaction. For example, ML systems are extremely sensitive to slight variations in their input data. Changes to an image’s orientation or lighting, or the addition of noise, can cause a model to generate alarmingly incorrect classifications or predictions \citep{nguyen2015deep}. Similar sensitivities apply with respect to changes or additions to a model’s training dataset. Adding new data to a model’s training set “can at times result in reduced overall accuracy, uncertain fairness outcomes, and reduced worst-subgroup performance” \citep[1]{shen2024data}. Even changing the order in which a training dataset is presented to a learning algorithm can result in substantial variations in the final model \citep{sogaard2023opacity}. Ultimately, even minor changes or updates to a model can have dramatic effects on the reasoning and behaviour of the system.

Model updating, therefore, is likely to result in intra-ML disagreements, or inconsistencies in the reasoning, performance, and behaviour of an ML systems as it evolves over time (diachronic evolution), along with inconsistencies in the reasoning, performance, and behaviour between two or more copies of an ML system deployed across multiple sites (synchronic variation) \citep{hatherley2023diachronic} (see Figure \ref{fig:fig1}).\footnote{While diachronic evolution can occur irrespective of the training data used in model updating, synchronic variation will only occur if each version of the model is updated separately using non-identical training datasets, or identical datasets presented to the learning algorithm in a different sequence \citep{hatherley2023diachronic}.}  To illustrate, consider the following two scenarios:

\begin{figure}
    \centering
    \includegraphics[scale=1]{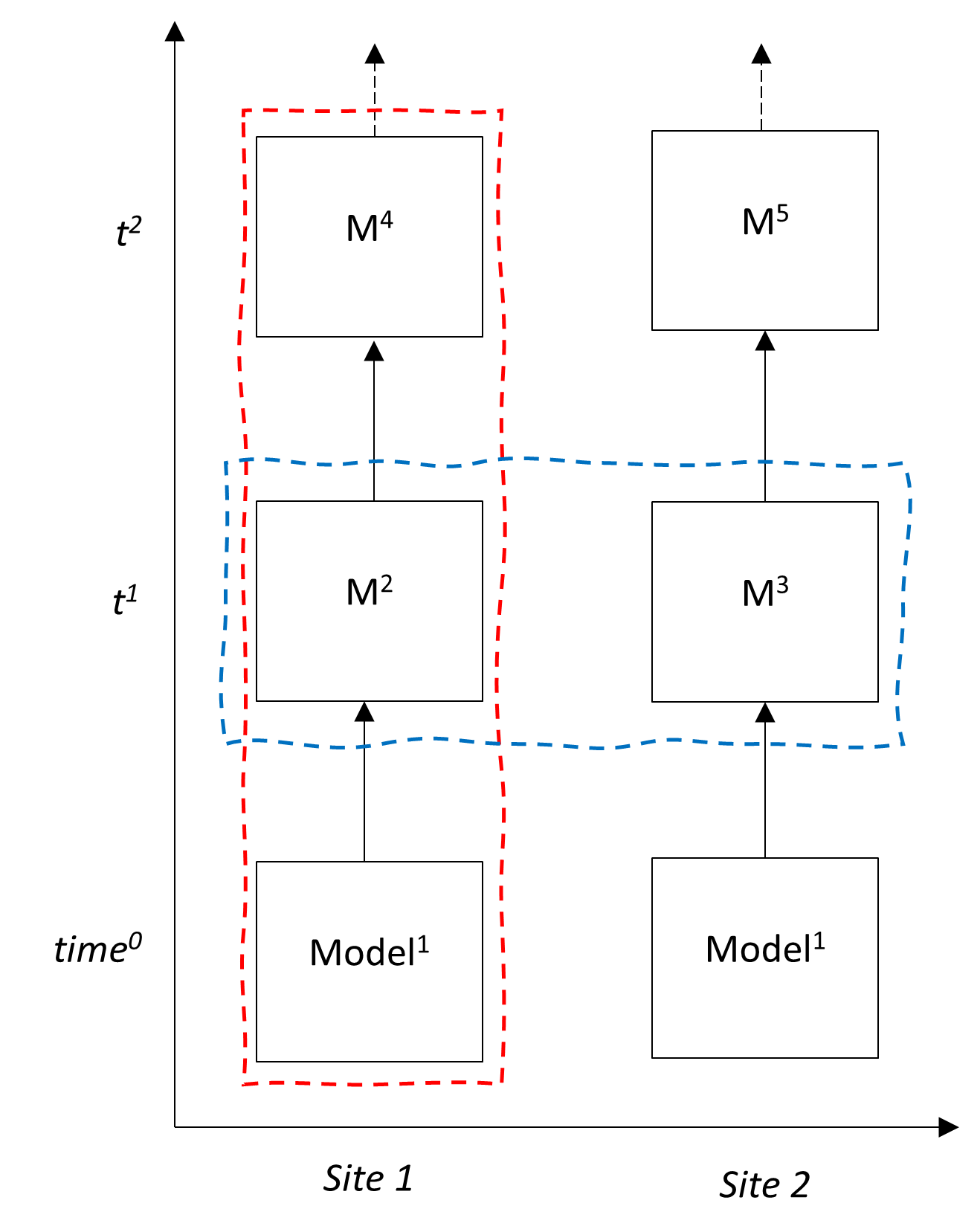}
    \caption{Schematic representation of diachronic evolution (red) and synchronic variation (blue) in a ML system deployed at two locations. Figure adapted from \cite{hatherley2023diachronic}.}
    \label{fig:fig1}
\end{figure}

\begin{enumerate}
    \item Diachronic evolution (see Figure \ref{fig:fig2}). Suppose a doctor uses a diagnostic ML system at time $t^1$. The doctor enters a series of inputs (\textit{a}, \textit{b}, and \textit{c}) into the model, which generates a diagnosis (\textit{x}) for a patient (P1). The doctor accepts the output and proceeds accordingly. Later, at time $t^2$, the system undergoes an update. The doctor then sees a different patient (P2) presenting with symptoms and a medical history that are identical in all relevant respects to P1. Again, the doctor inputs \textit{a}, \textit{b}, and \textit{c} into the model. But this time, the model generates a completely different diagnosis (\textit{y}).
        \begin{figure}
        \centering
        \includegraphics[scale=0.81]{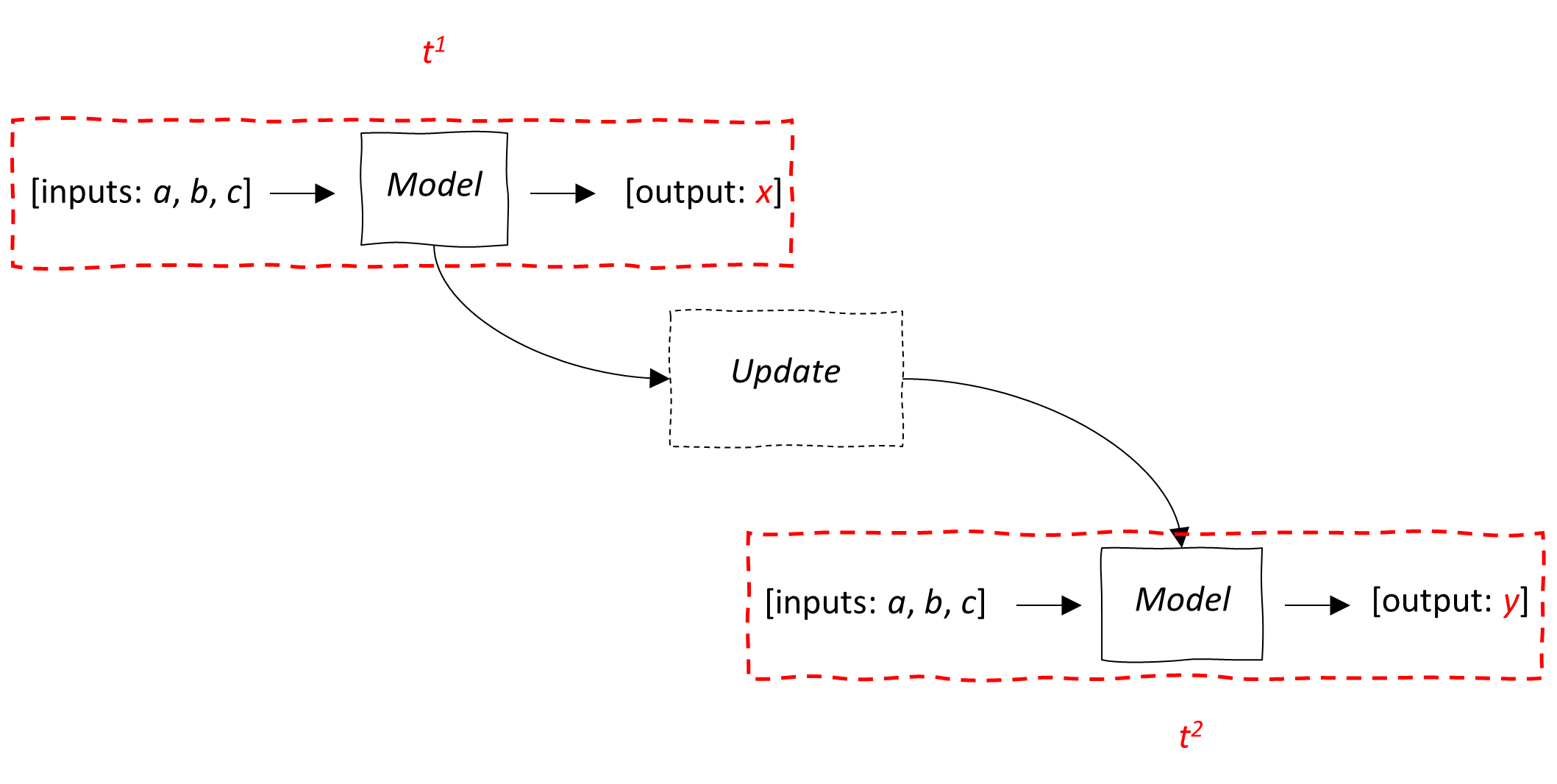}
        \caption{Schematic representation of intra-ML disagreement due to diachronic evolution.}
        \label{fig:fig2}
        \end{figure}
    \item Synchronic variation (see Figure \ref{fig:fig3}). Suppose that a doctor uses a diagnostic ML system at Hospital 1 at $t^1$. The doctor inputs \textit{a}, \textit{b}, and \textit{c} into the model, which generates diagnosis \textit{x} for P1. The doctor accepts the output and proceeds accordingly. The doctor then travels to Hospital 2 and uses the same system to assist in diagnosing P2, who presents with symptoms and a medical history that is identical in all relevant respects to P1. Both versions of the ML system, moreover, are updated separately, using local datasets that are specific to each clinical site, resulting in synchronic variation. Again, the doctor inputs \textit{a}, \textit{b}, and \textit{c} into the model at $t^1$. But this time, the model generates a different diagnosis, \textit{y}. 
\end{enumerate}

\begin{figure}
\centering
\includegraphics[scale=0.85]{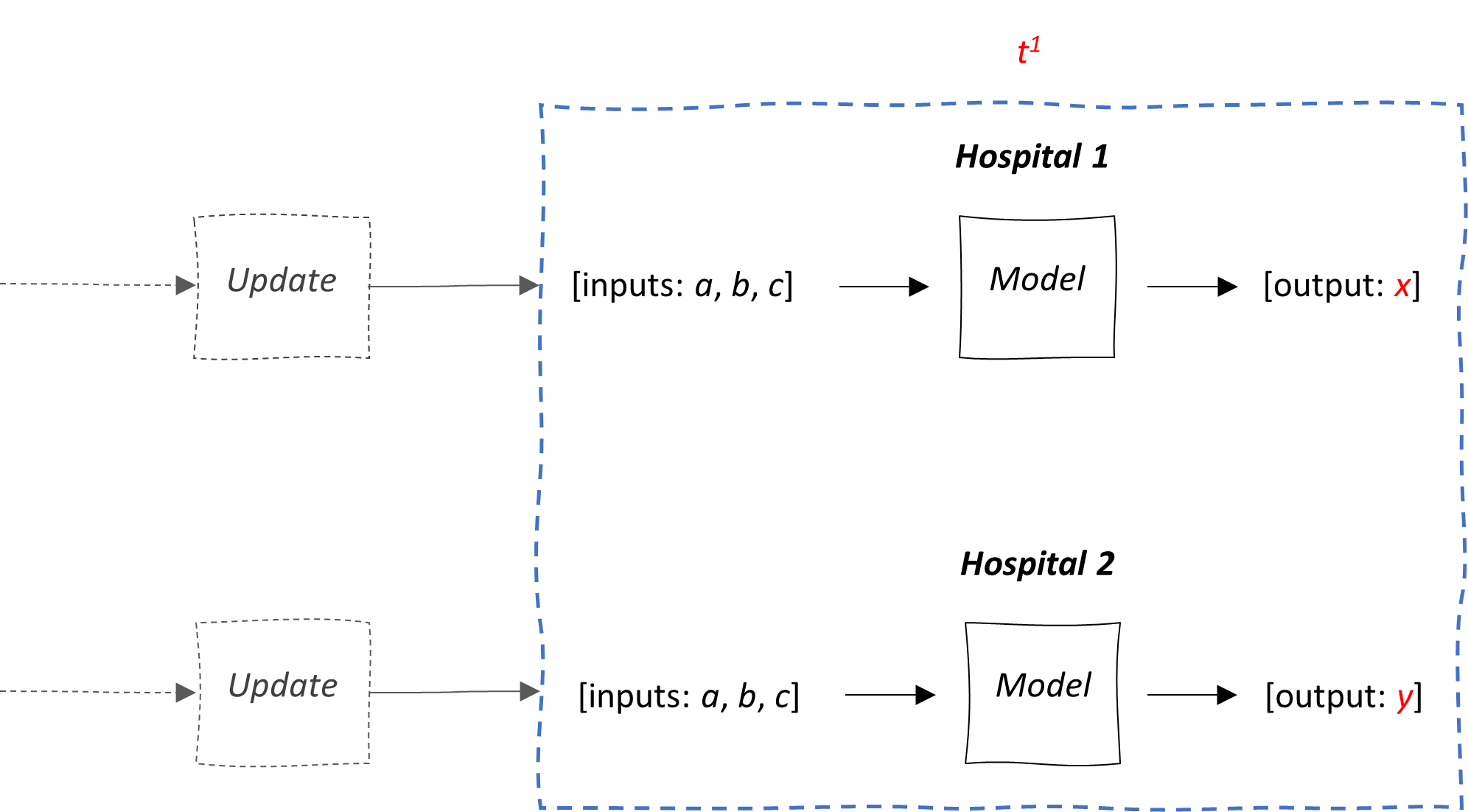}
\caption{Schematic representation of intra-ML disagreement due to synchronic variation.}
\label{fig:fig3}
\end{figure}

As discussed earlier, the mere fact that the current model is the most recent, up-to-date version of an ML system provides no straightforward guarantee that output \textit{y} is more correct than output \textit{x}. Users, therefore, may need to understand how and why the reasoning of the system has changed between updates to reliably and rationally resolve such disagreements, as I discuss further in section \ref{4}. 

Moreover, intra-ML disagreements and other unexpected changes in the behaviour, performance, or reasoning of an ML system can result in substantial drops in the performance of human-ML teams. Human users learn when it is appropriate to rely on the outputs of ML systems by developing mental models of these systems through repeated interactions with them \citep{bansal2019updates}. Mental models refer to “the internal, conceptual and operational representations that humans develop while interacting with complex systems” \citep[37]{jonassen1999mental}. However, model updates have been found to disrupt users’ mental models of ML systems, resulting in precipitous drops in the performance and accuracy of human-ML teams \citep{bansal2019updates}. While human-ML team performance eventually returns to its pre-update levels as human users update their mental models to align with the updated version of the system, such performance drops could present substantial health and safety risks during the immediate, post-update phase, and the subsequent recalibration phase in which users realign their mental models with the updated version of a system. 

\subsection{Summary and conclusion}\label{2.4}

Model updating can be used to reduce the risk of performance decline over time due to dataset shift. While the debate about how best to go about these updates is ongoing, at least three model updating frameworks have been proposed in the literature: conventional, dynamic, and continuous. While model updating can reduce the risk of datasets shift, it also presents epistemic and safety concerns. In particular, model updating is likely to result in instances of intra-ML disagreement, or inconsistencies in the reasoning, performance, and behaviour of a single ML system as it evolves over time. Unexpected changes or inconsistencies in an ML system’s behaviour such as these have been found to disrupt users’ mental models of these systems, resulting in substantial drops in the performance of human-ML teams.

\section{The black box problem}\label{3}

In this section, I leave model updating aside to discuss the so-called black box problem in ML. Section \ref{3.1} gives a brief overview of the problem. Section \ref{3.2} highlights a particular aspect of the black box problem, known as the reconciliation problem. Section \ref{3.3} considers the theory of computational reliabilism as a potential resolution to the reconciliation problem. Finally, section \ref{3.4} summarises and concludes section \ref{3}.

\subsection{The black box problem}\label{3.1}

It is by now common knowledge that ML systems, such as deep neural networks (DNNs), are opaque in the sense that human agents – including users, regulators, or the designers of the systems themselves – cannot understand key information about the system, or the system’s outputs, from the information that is in fact, or in principle, available to them \citep{burrell2016machine}. This characteristic of ML systems is thought to give rise to the black box problem, which “occurs whenever the reasons why an AI decision-maker has arrived at its decision are not currently understandable to [relevant stakeholders] because the system itself is not understandable to [them]” \citep[767]{wadden2022defining}.

The black box problem is complex and multifaceted. For example, opacity is thought to interfere with stakeholders’ ability to adequately assess the risks of ML systems ex ante \citep{braun2021primer}. Concerns have also been raised about users’ capacity to identify algorithmic biases or spurious reasoning in opaque ML systems \citep{challen2019artificial}. Others argue that users cannot be justified in accepting the outputs of an ML system whose reasoning they cannot understand \citep{shortliffe2018clinical}. The list goes on and on \citep{funer2022deception,hatherley2024data,muralidharan2024ai,sparrow2020high}. Indeed, some argue that the problem itself is a misnomer; there is no black box problem, only black box problems \citep{zednik2021solving}. 

For the purposes of this article, I focus on one specific aspect of the black box problem, known as the reconciliation problem.

\subsection{The reconciliation problem}\label{3.2}

The reconciliation problem in ML occurs when it is unclear how users ought to update their confidence in their own judgments in response to the outputs of ML systems whose reasoning they do not understand, or to reconcile these judgments with the outputs of such systems \citep{krishnan2020against}. It is particularly acute where the outputs of a model challenge or contradict the judgments of expert human users. For example, suppose a specialist doctor consults with a patient and concludes that treatment \textit{x} is most appropriate for their condition. The doctor then compares their recommendation with a medical DNN, which recommends the patient receive a conflicting treatment (\textit{y}). In such cases, it is not immediately clear how a doctor ought to proceed. Unlike human-human disagreements, the doctor cannot query the ML system to critically assess its reasoning. Nor can the doctor understand how the system arrived at treatment \textit{y} from the inputs it was given, or how the parameters of the system were induced from the data on which the system was trained. What results is an epistemic stalemate in which two agents with matching evidence and comparable expertise concerning a matter (M) disagree with respect to M \citep{grote2020ethics}. Philosophical epistemologists refer to such scenarios as instances of peer disagreement.

In the context of this article, the reconciliation problem is important since, by introducing intra-ML disagreement into ML-assisted decision-making scenarios, model updating exacerbates the core dilemma at the heart of this problem. Not only must users reconcile their own judgments with the outputs of an ML system; they must now also reconcile disagreements and inconsistencies within the ML system itself. The implications of this are discussed further in section \ref{4}.

\subsection{Computational reliabilism}\label{3.3}

One potential solution to the reconciliation problem comes from the theory of computational reliabilism. According to this theory, a user is justified in accepting the output of an ML system if the ML system is demonstrably reliable. In formal terms, computational reliabilism states that:

\begin{quote}
    if S’s believing \textit{p} at \textit{t} results from \textit{m}, then S’s belief in \textit{p} at \textit{t} is justified. 
    
    where S is a cognitive agent, \textit{p} is any truth‑valued proposition related to the results of a computer simulation, \textit{t} is any given time, and \textit{m} is a reliable computer simulation \citep[654]{duran2018grounds}.
\end{quote}

What does it mean for an ML system to be reliable? Key proponents of computational reliabilism suggest that an ML system’s reliability can be assessed according to three reliability indicators: (a) its “design, coding, execution, maintenance, and other technical features that make a system perform well (e.g., in terms of robustness, precision, and accuracy)”; (b) its use of “common knowledge, scientific theories, principles, hypotheses, and other relevant units of scientific analysis”; and (c) its alignment with “broader goals related to accepting AI and its outputs in diverse communities (e.g., scientific, academic, and public communities)” \citep[4]{duran2024understanding}.

Now, taking computational reliabilism seriously suggests that the reconciliation problem could simply be resolved by deferring to the judgement of whichever agent is more reliable \citep[see][]{duran2021afraid}. In the scenario from section \ref{3.1}, for instance, if the ML system is more reliable than the human doctor (or human doctors in general) at recommending treatments that improve patient health, then computational reliabilism suggests that the human doctor ought to defer to the ML system \citep[see][]{bjerring2021artificial}. 

The problem, however, is that ML systems will be used – and indeed, are already being used – in high-stakes decision-making scenarios in which it is unclear whether the system or the human user is in fact more reliable. While it is widely acknowledged that ML systems challenge the epistemic authority of human experts, most experts stop short of declaring the outright epistemic superiority of these systems \citep{funer2022deception,grote2020ethics,hatherley2020limits}. Moreover, while some studies have claimed to demonstrate that ML systems are in fact more reliable than human experts in certain tasks, critical scrutiny has typically revealed these claims to be exaggerated or misleading \citep{drogt2024ethical}.

One reason for this is that it is difficult to establish conclusively that a human expert, or human experts in general, are more reliable than an ML system, or vice versa. Reliability can be assessed at many different levels of abstraction, and comparing the reliability of two agents at each of these levels can result in different verdicts. For example, suppose we are comparing the reliability of human doctors and an ML system in correctly diagnosing a particular medical condition. The reliability of the human versus the ML system may be compared at the level of the population at large, at the level of the local community, at the level of protected patient groups, at the level of patients with a particular disease, and so on. Comparing the reliability of the human and ML system at each of these levels at each of these levels may yield different results, and there may be no obvious reason to value one specific level above all the others.

Currently, therefore, computational reliabilism is insufficient to resolve the reconciliation problem.

\subsection{Summary and conclusion}\label{3.4}

ML systems are opaque in the sense that human agents cannot understand key information about the system, or the system’s outputs, from the information that is in fact, or in principle, available to them. This characteristic of ML systems gives rise to the black box problem, in which stakeholders cannot understand how a system arrives at each of its outputs. This problem generates a variety of concerns in high-stakes decision-making environments. One such concern is the reconciliation problem, which occurs when it is unclear how users ought to reconcile their own judgements with the outputs of ML systems whose reasoning they do not understand. Computational reliabilism provides a potential solution to this problem. However, it only succeeds in decision-making scenario where an ML system is demonstrably more reliable than a human expert, or human experts in general (or vice versa). Currently, however, such scenarios both occur rarely (if at all) and are difficult to conclusively establish, yet the use of ML systems in high-stakes decision-making contexts persists. Currently, therefore, computational reliabilism does not provide an adequate solution to the reconciliation problem.

\section{The problem of update opacity}\label{4}

In this section, I outline the problem of update opacity in ML. Section \ref{4.1} provides a definition of update opacity. Section \ref{4.2} provides an account of the problem of update opacity and the distinctive challenges it presents. Section \ref{4.3} addresses a potential double standard objection to the problem of update opacity. Finally, section \ref{4.4} summarises and concludes section \ref{4}.

\subsection{What is update opacity?}\label{4.1}

Update opacity can be understood in contrast to other types of opacity in ML. As noted above, ML systems are often opaque in the sense that human agents cannot understand key information about the system, or the system’s outputs, from the information that is in fact, or in principle, available to them. More specifically, however, DNNs exhibit at least two different types of opacity: 

\begin{itemize}
    \item \textit{Training opacity} occurs “when expert humans cannot, upon inspection, say how, in general, the parameters of the DNN were induced as a result of its training data” \citep[225]{sogaard2023opacity}.
    \item \textit{Inference opacity} occurs “when expert humans cannot, upon inspection, say why, in general, [a model] predicts an output y as a result of an input x” \citep[225]{sogaard2023opacity}. 
\end{itemize}

DNNs exhibit training opacity due to fact that learning algorithms are highly sensitive to even minor variations in a training dataset (e.g. the order in which this data is presented to the learning algorithm, discussed earlier in section \ref{2.3}). As a result, users cannot understand or anticipate how an ML system arrives at its final state from the data used to train it. In contrast, inference opacity occurs when the size of a model (i.e. the number of parameters it contains) exceeds the limits of human short-term memory. Inspecting a diagnostic DNN, for example, would provide a doctor with no insight into the model’s reasoning since these models often contain millions of weighted parameters attached to specific textual or visual features that the model has learned to identify during training.

Update opacity can be understood as a subtype of inference opacity. I define it as follows:

\begin{itemize}
    \item \textit{Update opacity} occurs when expert humans cannot, upon inspection, say why, in general, the current version of a model predicts an output (\textit{y}) as a result of an input (\textit{x}), while a previous version of the model predicts a different output (\textit{z}), as a result of \textit{x}. 
\end{itemize}

Like inference opacity, update opacity interferes with users’ ability to understand how a model arrives at a particular output from a certain set of inputs. However, inference opacity is static and atemporal insofar as it concerns users’ understanding of the current version ML system at a specific point in time, independent of its previous iterations. In contrast, update opacity is dynamic and temporal insofar as it concerns users’ understanding the differences and relations between two or more distinct versions of the same model at different points in time.

\subsection{Implications}\label{4.2}

The problem of update opacity is most acute in situations where a user is already familiar with, and confident in, a system’s performance pre-update. In such cases, the user is likely to hold certain expectations with respect to the system’s behaviour, and to have developed a functional mental model of when to rely on the system’s outputs and when to reject them. Once the model is updated, however, these expectations may be undermined, and the user’s mental model may become outdated. In such cases, update opacity could present a variety challenges in ML-assisted decision-making over and above existing concerns associated with the black box problem in ML more generally. 

For example, update opacity may interfere with users’ ability to resolve instances of intra-ML disagreement. As discussed in section \ref{2.3}, model updating is likely to introduce inconsistencies in the reasoning and behaviour of a single ML system as it evolves over time. Moreover, model updating does not guarantee better outputs across all decision-making scenarios, resulting in situations in which users must decide whether to defer to the output of the previous system, or trust the new output of the updated system. However, update opacity compromises users’ capacity to interrogate changes in the reasoning of a system to determine whether inconsistencies in its outputs between updates are the result of performance improvement or degradation. The additional decision-making complexity generated by intra-ML disagreements may also exacerbate the risk of cognitive overload. Users may feel inclined to defer to the system out of mental exhaustion, or under the assumption that, because the system is the most up-to-date version of the model, its output must therefore be superior to its predecessor. 

Update opacity may also increase the time it takes for users to recalibrate their mental models of an ML system to adequately reflect the changes that have occurred in the system between updates, resulting in downstream safety risks. As discussed in section \ref{2.3}, model updating has been found to disrupt users’ mental models of ML systems, resulting in substantial performance drops post-update. As users gradually update their outdated mental model of a system to align with the updated version of the model, the performance of the human-ML team eventually returns to pre-update levels. However, where users do not understand how a model’s reasoning has changed between updates, their capacity to adapt to these changes may be slowed, resulting in longer latency periods. Update opacity, therefore, may interfere with users’ capacity to maintain reliable and up-to-date mental models of ML systems that undergo model updates, potentially resulting in avoidable decision-making errors or safety incidents.

\subsection{A double standard?}\label{4.3}

The problem of update opacity may be accused of exhibiting a double standard. After all, human experts often update their knowledge or practices, and in many cases, may do so without informing their colleagues (or at least without their colleagues understanding precisely how the expert’s knowledge and practices may have changed). Indeed, many readers may be familiar with the experience of asking an authority for advice on a matter, then returning later for advice on a similar matter only to receive a completely different, if not contradictory response. Human experts, therefore, could be understood as exhibiting their own form of update opacity. 

Moreover, decision-making guidelines are regularly updated in high-stakes domains such as medicine. For example, clinical practice guidelines for treating diseases like diabetes or cancer are revised frequently to incorporate the latest research findings and ensure that healthcare professionals are using the most effective and evidence-based approaches to patient care. However, the reasons underlying these changes may be unknown to doctors who use them to make decisions. Why, then, should we be more concerned about the problem of update opacity in ML than we are about the problem of update opacity as it applies to human beings or decision-making guidelines?

However, the objection is misguided. One reason for this is that update opacity can generate problems no matter where it occurs. At the beginning of the COVID-19 pandemic, for example, recommendations for the use of personal protective equipment in healthcare settings changed frequently and without clear reason, leading to significant confusion and distrust amongst doctors \citep{hoernke2021frontline}.\footnote{Thanks to an anonymous reviewer for this example.}  In general, however, we ought to be more concerned about updating and update opacity in ML systems because they have several distinctive characteristics. Unlike ML systems, for example, human experts can answer questions about why they are responding differently to identical decision-making scenarios at different times. Such explanations can greatly reduce the apparent update opacity of human experts in comparison to ML systems. 

In addition, human experts are more robust to changes and updates in their knowledge and practices than ML systems. As discussed above, even minor updates to an ML system can result in dramatic changes to their responses to certain input data. The rationale behind such changes may be strong or weak, yet human users are unable to scrutinise these rationales to find out for themselves. While human experts may give differing advice over time, these differences could be attributed to minor contextual changes that may be justified or explained. The inconsistencies of ML systems, on the other hand, may occur for reasons that are much harder to discern. This is particularly concerning because users of ML systems are prone to automation bias, or “omission and commission errors resulting from the use of automated cues as a heuristic replacement for vigilant information seeking and processing” \citep[47]{mosier2017automation}. Model updating may exacerbate such biases, and update opacity may interfere with users’ capacity to correct themselves. Finally, human experts typically operate at a smaller scale, impacting fewer decisions than ML systems, which can influence millions of decisions simultaneously in high-stakes contexts. Update opacity in ML is thus magnified in its potential consequences.

\subsection{Summary and conclusion}\label{4.4}

Update opacity occurs when when expert humans cannot, upon inspection, say why, in general, the current version of a model predicts an output \textit{y} as a result of an input \textit{x}, while the previous version of the model predicts a different output \textit{z} as a result of \textit{x}. The problem of update opacity is most acute in situations where a user is already familiar with, and confident in, a system’s performance pre-update. In such cases, update opacity presents some distinctive epistemic and safety risks. For example, update opacity is likely to exacerbate performance drops in human-ML teams that result from model updating, and to increase the complexity of reconciling human judgments with algorithmically generated outputs. One may object that the problem of update opacity exhibits a double standard since human experts can also be understood as exhibiting a form of update opacity. However, this objection neglects several relevant differences between human experts and ML systems. 

\section{What can be done?}\label{5}

In this section, I evaluate a range of strategies that could be used to address the problem of update opacity in ML going forward. Section \ref{5.1} gives an analysis of “conventional” strategies, by which I mean strategies that are already being used to address the black box problem in ML more generally. Section \ref{5.2} proposes and assess several alternative strategies, targeted specifically to the problem of update opacity. Section \ref{5.3} considers the implications of the preceding analysis for the prioritisation of “interpretable” AI systems over their black box counterparts in high-stakes decision-making. Section \ref{5.4} summarises and concludes section \ref{5}.

\subsection{Conventional strategies}\label{5.1}

A variety of strategies have been proposed to address the black box problem in ML over the past several years. Here I consider three such “conventional” strategies: post-hoc explanation models, model reporting, and performance validation. 

\subsubsection{Post-hoc explanations}\label{5.1.1}

Post-hoc explanation models refer to a suite of ML systems that reverse-engineer explanations for the inner workings and reasoning processes of other, black box ML systems by perturbing the inputs of these systems and analysing the effects of such perturbations on the system’s outputs. A broad range of post-hoc explanation strategies have been developed over the past several years, including activation maximisation, features importance, prototype selection, and so on \citep{guidotti2018survey,minh2022explainable}. Saliency maps, for example – one of the most popular types of post-hoc explanation models \citep{borys2023explainable} – produce visual overlays for inputted images that highlight the areas of an image that were most influential in determining a system’s output. Where the highlighted regions correspond with human-interpretable features (e.g. lesions or tumours), these explanations can assist users in answering certain “why-questions” about a black box system’s output (e.g. why did the system generate this specific output?) \citep{zednik2021solving}. 

A problem, however, is that post-hoc explanation models cannot replicate, but only approximate, the exact reasoning that an ML system has taken to arrive at an output \citep{babic2021beware,babic2023algorithmic,ghassemi2021false}. This is concerning, since it means that such models may generate post-hoc rationalisations for the outputs of black box ML systems, rather than genuine explanations. This may increase the likelihood that users will accept incorrect or dangerous outputs. Indeed, empirical studies into the effects of post-hoc explanation models have found that these models can generate intuitively reassuring explanations for even completely untrained black box models (despite how implausible the outputs of these systems may be) \citep{adebayo2018sanity}. They have also been found to promote overconfidence amongst users in the outputs of black box systems due to information overload \citep{poursabzi2021manipulating}. In some cases, affixing a post-hoc explanation model to a black box system results in worse overall performance than just using the black box system on its own \citep{jesus2021can}. 

With respect to the problem of update opacity, moreover, post-hoc explanations may reduce the inference opacity of ML systems, but they cannot reduce the update opacity of these systems. This is because post-hoc explanations, like inference opacity, are static and atemporal; they are limited to providing explanations for the current version of an ML system, independent of its previous iterations. As a result, post-hoc explanations cannot provide any insight into how a system’s reasoning processes or behaviours may have changed since its last update – only how the system reasons now.

\subsubsection{Model reporting}\label{5.1.2}

Model reporting involves communicating information to users about the characteristics, limitations, and design features of black box system to promote their safe and responsible use. The idea is that such information can help users to appropriately rely on these systems, even where their inner reasoning processes and operations cannot be understood \citep{kroll2018fallacy}. Model cards are one example of model reporting, which refer to “short documents accompanying trained machine learning models that provide benchmarked evaluation in a variety of conditions, such as across different cultural, demographic, or phenotypic groups (e.g., race, geographic location, sex, Fitzpatrick skin type) and intersectional groups (e.g., age and race, or sex and Fitzpatrick skin type) that are relevant to the intended application domains” \citep[220]{mitchell2019model}. In addition, some researchers suggest that users ought to be provided with further information about an ML system’s subjective “point-of-view” (e.g. whether a system tends to diagnose certain medical conditions more liberally or more conservatively), and how this point-of-view compares to the user’s own \citep{cai2019hello}.

However, while model reporting strategies are useful in providing users with information about an ML system in general, it is unclear how such reporting strategies could address the challenges associated with update opacity. Developers often neglect to revise their documentation as models are updated over time (Wang et al., 2022), and even where such documentation is updated, the revisions may not be detailed enough to assist users in anticipating or appropriately respond to changes in the behaviour of an updated ML system. For example, some model cards may contain comparative quantitative analyses of different versions of a model \citep[see the model card for "toxicity in text" in][]{mitchell2019model}. But such comparative information is likely to be too broad to assist users in specific decision-making scenarios. While model reporting may assist in addressing the problem of update opacity, therefore, it is currently an incomplete solution. 

\subsubsection{Performance validation}\label{5.1.3}

Performance validation involves rigorously testing the performance of ML systems to ensure their reliability and robustness in high-stakes decision-making scenarios. With respect to model updating, performance validation involves rigorously testing an updated ML system to ensure that its performance has generally improved as a result of the update. Echoing the discussion of computational reliabilism (discussed in section \ref{3.3}), the idea here is that update opacity in an ML system may not necessarily a problem so long as users know that the system is robust and reliable \citep{ghassemi2021false}. 

However, improved model performance does not guarantee improved outcomes \citep{hatherley2024virtues}. What matters most is the performance of the human-ML team, not the performance of the ML system in isolation. As discussed previously, model updating and performance validation cannot ensure that unexpected changes to the reasoning or behaviour of an ML system will not occur as a result of model updating, and such changes can significantly compromise the performance of human-ML teams. While performance validation is essential to safe and effective ML-assisted decision-making, therefore, it does not address the problem of update opacity. 

\subsection{Alternative strategies}\label{5.2}

Several alternative strategies could also be developed that are more specifically tailored to addressing the problem of update opacity. Here I consider three such strategies: bi-factual explanations, dynamic model reporting, and update compatibility.

\subsubsection{Bi-factual explanations}\label{5.2.1}

Contrastive questions are those that ask, “Why \textit{P} rather than \textit{Q}?” (Miller 2020). Two types of contrastive questions are distinguished in the literature. Counterfactual questions are those in which \textit{Q} refers to a counterfactual case, or “foil,” that did not actually occur. For example, “Why did the ML system diagnose this patient with condition \textit{x} rather than condition \textit{y}?” In contrast, bi-factual questions query “why some fact happened in one situation while another fact, called the surrogate, happened in another (presumably similar) situation” \citep[8]{miller2021contrastive}. Unlike counterfactual questions, both \textit{P} and \textit{Q} in a bi-factual question refer to actual cases that did in fact occur. Because of this, bi-factual questions can target the problem of update opacity directly, while counterfactual questions cannot. The following question, for example, is bi-factual: “Why did the current version of the ML system diagnose Patient 1 with condition \textit{y} from input data \textit{a}, \textit{b}, and \textit{c}, while a previous version of the system diagnosed a similar patient, Patient 2, with condition \textit{z} from \textit{a}, \textit{b}, and \textit{c}?” 

Counterfactual explanations have become extremely popular approaches to post-hoc explanation in recent years. These explanations provide an account of how the input data would need to change to generate \textit{Q} rather than \textit{P}. For example, “You were denied a loan because your annual income was £30,000. If your income had been £45,000, you would have been offered a loan” \citep[844]{wachter2017counterfactual}. However, little attention has thus far been paid to the development of post-hoc explanation models that provide bi-factual explanations \citep{miller2021contrastive}. This is a problem since, while counterfactual explanations can assist in reducing inference opacity, they cannot assist in reducing update opacity. Further research into the development of post-hoc bi-factual explanations, therefore, could significantly assist in addressing the problem of update opacity going forward.

A problem, however, is that bi-factual explanations are likely to exhibit the same weaknesses as post-hoc explanation more generally: they will only provide approximations of any changes in reasoning between different versions of a model, rather than explain the actual changes that occurred. Again, such post-hoc rationalisations may increase the risk of overconfidence and overreliance on incorrect or dangerous outputs. Moreover, they also risk adding to the cognitive burdens of human decision-makers, thereby increasing their risk of cognitive overload. 

\subsubsection{Dynamic model reporting}\label{5.2.2}

As discussed earlier, a limitation of model reporting strategies is that updates to the information provided are often either shallow or infrequent. This limitation could be addressed through dynamic model reporting, in which users are informed about changes that have occurred to the model after each update. For example, developers could provide users with update cards that describe updated performance metrics, notable shifts in the demographic constitution of training datasets, changes to the potential limitations and error boundaries of the model, changes in the potential biases of the model, changes to the system’s subjective “point-of-view,” and so on. Alternatively, developers could update existing model cards with such changes, highlighting the information contained in the model card that has changed post-update. 

But the benefits of this approach are likely to be limited according to how frequently a model is updated. If a system is updated every other day, for example, maintaining update cards is likely to be practically infeasible.\footnote{Thanks to an anonymous reviewer for pointing this out.}  Moreover, it is unclear how useful such information would be to users trying to interpret behavioural changes in an updated system. The provision of such information may not reduce the epistemic or safety risks of update opacity since it does not appear to address this type of opacity directly. Update cards would only inform users about differences in the general characteristics of the model itself, rather than the differences in reasoning between different versions of a model. Like bi-factual explanations, dynamic model reporting may also cause more harm than good by adding to the cognitive burdens of human decision-makers and increasing their risk of cognitive overload.

\subsubsection{Update compatibility}\label{5.2.3}

Developers could also minimise the problem of update opacity by maximising compatibility scores between updates. The higher the compatibility score, the greater the consistency in correct outputs between different versions of a model. In formal terms: “The compatibility score \textit{C} of an update $h_{2}$ to $h_{1}$ is given by the fraction of examples of which $h_{1}$ recommends the correct action, $h_{2}$ also recommends the correct action” \citep[2431]{bansal2019updates}. Higher compatibility between updates could reduce disruptions to the mental models of users, and their associated safety risks, by reducing the likelihood of unexpected changes in the behaviour and outputs of an ML system. 

However, there is a trade-off between compatibility and performance. Higher compatibility scores tend to reduce the overall performance improvements that are made to a model between updates \citep{bansal2019updates}. Higher compatibility, therefore, could limit the degree to which model updating addresses the risks associated with dataset shift. Moreover, ground truths may not always be available to assess compatibility in certain decision-making contexts. In medicine, for example, there is no universally agreed-upon definition or diagnostic test for sepsis. This lack of ground truth is likely to make it challenging to reliably assess the compatibility of updates to ML systems, potentially limiting the usefulness of compatibility scores and complicating efforts to balance system performance with continuous improvement.\footnote{Thanks to an anonymous reviewer for pointing this out.}

\subsection{Interpretable AI}\label{5.3}

Recently, several prominent researchers have argued that interpretable AI systems should be prioritised over their black box counterparts in high-stakes decision-making environments, absent clear and compelling justification \citep{babic2023algorithmic,rudin2019stop}. Interpretable AI refers to systems whose complexity has been restricted during pre-training to ensure that users can understand how the resulting model arrive at specific outputs from certain inputs \citep{molnar2020interpretable}. Decision trees, for instance, are a paradigmatic example of interpretable AI (although they, too, can become opaque if the model becomes too large). But constraints can also be applied during the development of otherwise opaque ML systems to ensure their interpretability \citep{rudin2019we}. A virtue of prioritising interpretable AI over black box systems in high-stakes environments is that, not only could it eliminate update opacity and reduce the latency of users’ mental models with respect to adapting to updated versions of an ML system; it would also address all the other issues associated with the black box problem in ML.

However, prioritising interpretable systems over their opaque counterparts is often criticised for its potential to result in worse outcomes \citep{london2019artificial}. It is often accepted that there is a trade-off between accuracy and interpretability in AI systems; the more accurate a system, the less interpretable it tends to be (and vice versa). As a result, prioritising the use of interpretable systems appears likely to involve using less accurate decision aids in high-stakes scenarios. 

But the reality of the accuracy-interpretable trade-off is heavily debated. In many cases, interpretable systems appear to perform equally well as their black box counterparts \citep{rudin2019stop}. Moreover, superior accuracy in an ML system does not guarantee superior outcomes \citep{hatherley2024virtues}. As mentioned previously, what matters most is the performance of the human-ML team, not the performance of the ML system in isolation. Whether using a system that is interpretable, but less accurate produces worse outcomes than using a system that is opaque, but more accurate is an empirical matter that cannot be settled \textit{a priori}. Given the potential negative impact of update opacity on the performance of clinician-AI teams, sacrificing technical performance to overcome opacity may in fact result in better human-AI team performance overall.

\subsection{Summary and conclusion}\label{5.4}

A variety of conventional strategies may be deployed to address the problem of update opacity, including post-hoc explanation models, model reporting, and performance validation. However, each of these strategies presents its own risks or carries limitations that prevent them from providing completely satisfactory solutions. Several alternative strategies may be developed that are tailored to address the problem of update opacity more directly, including bi-factual explanations, dynamic model reporting, and update compatibility. Again, however, these strategies carry limitations or trade-offs that prevent them from being completely satisfactory. Until such limitations are addressed or superior strategies are developed, prioritising interpretable AI systems over their black box counterparts in high-stakes environments may often be the most optimal solution. 

\section{Conclusion}\label{6}

ML systems are vulnerable to performance decline over time due to dataset shift. Regular model updating is commonly suggested for addressing these performance issues. However, model updating presents a variety of epistemic and safety issues requiring careful consideration and analysis. Model updating also introduces a new sub-type of inference opacity into AI-assisted decision-making – update opacity – which occurs when expert humans cannot, upon inspection, say why, in general, the current version of a model predicts an output (\textit{y}) as a result of an input (\textit{x}), while the previous version of the model predicts a different output (\textit{z}), as a result of \textit{x}. Update opacity interferes with users’ capacity to understand changes in the reasoning and behaviour of ML systems as they evolve over time.  This could present downstream safety risks by adding to the complexity and cognitive burden of AI-assisted decision-making, and by increasing the time it takes for users recalibrate their mental models of an updated ML system.

A variety of strategies have been developed to address the black box problem more generally, including explainability methods, model reporting, and performance validation. But these strategies are poorly equipped to address the problem of update opacity. Several alternative strategies could be developed or pursued that target the problem of update opacity more directly, including bi-factual explanations, dynamic updating, and update compatibility. However, each of these strategies presents its own risks or carries substantial limitations that prevent them from being entirely satisfactory. Until such limitations are addressed or superior strategies are developed, prioritising interpretable AI systems over their black box counterparts in high-stakes environments may often be the most optimal solution.

Further research is needed to address the epistemic and safety concerns associated with model updating and update opacity. How can trust among users be maintained when model updating and update opacity may make it less clear over time how outputs have been reached? What methods can be employed to train users to adapt to changes in a model’s behaviour post-update, despite its update opacity? How might a model’s explainability be maintained when significant changes have been made during the updating process? My hope is that this article serves as a useful starting point for understanding and addressing these challenges going forward.

\bigskip

\backmatter

\bmhead{Acknowledgements} 

Thanks to Jens Christian Bjerring, Thomas Grote, Karin Jongsma, Lauritz Munch, Giorgia Pozzi, and two anonymous referees for written comments on earlier versions of this article that greatly improved the final product. Thanks also to the audience at Aarhus University’s Workshop on Opacity, Explainability, and Reliability in AI and the audience at UMC Utrecht’s AI Ethics Seminar Series for critical commentary and discussion.

\bmhead{Funding declaration}

The research for this article was supported by a Carlsbergfondet Young Researcher Fellowship (CF20-0257).

\bigskip

\bibliography{references}

\begin{thebibliography}{56}
\providecommand{\natexlab}[1]{#1}
\providecommand{\doi}[1]{\url{https://doi.org/#1}}
\providecommand{\url}[1]{\texttt{#1}}
\providecommand{\urlprefix}{}

\bibitem[{Adam et~al.(2022)Adam, George Alexandru and Chang, Chun-Hao Kingsley and Haibe-Kains, Benjamin and Goldenberg, Anna}]{adam2022error}
Adam GA, Chang CHK, Haibe-Kains B, Goldenberg A.
\newblock Error amplification when updating deployed machine learning models.
\newblock Proceedings of Machine Learning Research. 2022;182:715--740.

\bibitem[{Adebayo et~al.(2018)Adebayo, Julius and Gilmer, Justin and Muelly, Michael and Goodfellow, Ian and Hardt, Moritz and Kim, Been}]{adebayo2018sanity}
Adebayo J, Gilmer J, Muelly M, Goodfellow I, Hardt M, Kim B.
\newblock Sanity checks for saliency maps.
\newblock Advances in Neural Information Processing Systems. 2018;31:1--11.

\bibitem[{Babic and Cohen(2023)Babic, Boris and Cohen, I Glenn}]{babic2023algorithmic}
Babic B, Cohen IG.
\newblock The algorithmic explainability "bait and switch".
\newblock Minnesota Law Review. 2023;108:857--909.

\bibitem[{Babic et~al.(2021)Babic, Boris and Gerke, Sara and Evgeniou, Theodoros and Cohen, I Glenn}]{babic2021beware}
Babic B, Gerke S, Evgeniou T, Cohen IG.
\newblock Beware explanations from AI in health care.
\newblock Science. 2021;373(6552):284--286.

\bibitem[{Bansal et~al.(2019)Bansal, Gagan and Nushi, Besmira and Kamar, Ece and Weld, Daniel S and Lasecki, Walter S and Horvitz, Eric}]{bansal2019updates}
Bansal G, Nushi B, Kamar E, Weld DS, Lasecki WS, Horvitz E.
\newblock Updates in human-AI teams: Understanding and addressing the performance/compatibility tradeoff.
\newblock Proceedings of the AAAI Conference on Artificial Intelligence. 2019;33(1):2429--2437.

\bibitem[{Bjerring and Busch(2021)Bjerring, Jens Christian and Busch, Jacob}]{bjerring2021artificial}
Bjerring JC, Busch J.
\newblock Artificial intelligence and patient-centered decision-making.
\newblock Philosophy \& Technology. 2021;34:349--371.

\bibitem[{Borys et~al.(2023)Borys, Katarzyna and Schmitt, Yasmin Alyssa and Nauta, Meike and Seifert, Christin and Kr{\"a}mer, Nicole and Friedrich, Christoph M and Nensa, Felix}]{borys2023explainable}
Borys K, Schmitt YA, Nauta M, Seifert C, Kr{\"a}mer N, Friedrich CM, et~al.
\newblock Explainable AI in medical imaging: An overview for clinical practitioners -- Beyond saliency-based XAI approaches.
\newblock European Journal of Radiology. 2023;162:110786.

\bibitem[{Braun et~al.(2021)Braun, Matthias and Hummel, Patrik and Beck, Susanne and Dabrock, Peter}]{braun2021primer}
Braun M, Hummel P, Beck S, Dabrock P.
\newblock Primer on an ethics of AI-based decision support systems in the clinic.
\newblock Journal of Medical Ethics. 2021;47(12):e3.

\bibitem[{Burrell(2016)Burrell, Jenna}]{burrell2016machine}
Burrell J.
\newblock How the machine ‘thinks’: Understanding opacity in machine learning algorithms.
\newblock Big Data \& Society. 2016;3(1):2053951715622512.

\bibitem[{Cai et~al.(2019)Cai, Carrie J and Winter, Samantha and Steiner, David and Wilcox, Lauren and Terry, Michael}]{cai2019hello}
Cai CJ, Winter S, Steiner D, Wilcox L, Terry M.
\newblock "Hello AI": Uncovering the onboarding needs of medical practitioners for human-AI collaborative decision-making.
\newblock Proceedings of the ACM on Human-Computer Interaction. 2019;3(CSCW):1--24.

\bibitem[{Challen et~al.(2019)Challen, Robert and Denny, Joshua and Pitt, Martin and Gompels, Luke and Edwards, Tom and Tsaneva-Atanasova, Krasimira}]{challen2019artificial}
Challen R, Denny J, Pitt M, Gompels L, Edwards T, Tsaneva-Atanasova K.
\newblock Artificial intelligence, bias and clinical safety.
\newblock BMJ Quality \& Safety. 2019;28(3):231--237.

\bibitem[{Drogt et~al.(2024)Drogt, Jojanneke and Milota, Megan and van den Brink, Anne and Jongsma, Karin}]{drogt2024ethical}
Drogt J, Milota M, van~den Brink A, Jongsma K.
\newblock Ethical guidance for reporting and evaluating claims of AI outperforming human doctors.
\newblock {npj} Digital Medicine. 2024;7(1):271.

\bibitem[{Dur{\'a}n and Formanek(2018)Dur{\'a}n, Juan M and Formanek, Nico}]{duran2018grounds}
Dur{\'a}n JM, Formanek N.
\newblock Grounds for trust: Essential epistemic opacity and computational reliabilism.
\newblock Minds and Machines. 2018;28:645--666.

\bibitem[{Dur{\'a}n et~al.(2024)Dur{\'a}n, Juan M and van der Vloed, David and Ruifrok, Arnout and Ypma, Rolf JF}]{duran2024understanding}
Dur{\'a}n JM, van~der Vloed D, Ruifrok A, Ypma RJ.
\newblock From understanding to justifying: Computational reliabilism for AI-based forensic evidence evaluation.
\newblock Forensic Science International: Synergy. 2024;9:100554.

\bibitem[{Dur{\'a}n and Jongsma(2021)Dur{\'a}n, Juan Manuel and Jongsma, Karin Rolanda}]{duran2021afraid}
Dur{\'a}n JM, Jongsma KR.
\newblock Who is afraid of black box algorithms? On the epistemological and ethical basis of trust in medical AI.
\newblock Journal of Medical Ethics. 2021;47(5):329--335.

\bibitem[{Feng et~al.(2022)Feng, Jean and Phillips, Rachael V and Malenica, Ivana and Bishara, Andrew and Hubbard, Alan E and Celi, Leo A and Pirracchio, Romain}]{feng2022clinical}
Feng J, Phillips RV, Malenica I, Bishara A, Hubbard AE, Celi LA, et~al.
\newblock Clinical artificial intelligence quality improvement: Towards continual monitoring and updating of AI algorithms in healthcare.
\newblock {npj} Digital Medicine. 2022;5(1):66.

\bibitem[{Finlayson et~al.(2021)Finlayson, Samuel G and Subbaswamy, Adarsh and Singh, Karandeep and Bowers, John and Kupke, Annabel and Zittrain, Jonathan and Kohane, Isaac S and Saria, Suchi}]{finlayson2021clinician}
Finlayson SG, Subbaswamy A, Singh K, Bowers J, Kupke A, Zittrain J, et~al.
\newblock The clinician and dataset shift in artificial intelligence.
\newblock New England Journal of Medicine. 2021;385(3):283--286.

\bibitem[{Funer(2022)Funer, Florian}]{funer2022deception}
Funer F.
\newblock The deception of certainty: How non-interpretable machine learning outcomes challenge the epistemic authority of physicians. A deliberative-relational approach.
\newblock Medicine, Health Care and Philosophy. 2022;25(2):167--178.

\bibitem[{Ghassemi et~al.(2021)Ghassemi, Marzyeh and Oakden-Rayner, Luke and Beam, Andrew L}]{ghassemi2021false}
Ghassemi M, Oakden-Rayner L, Beam AL.
\newblock The false hope of current approaches to explainable artificial intelligence in health care.
\newblock The Lancet Digital Health. 2021;3(11):e745--e750.

\bibitem[{Grote and Berens(2020)Grote, Thomas and Berens, Philipp}]{grote2020ethics}
Grote T, Berens P.
\newblock On the ethics of algorithmic decision-making in healthcare.
\newblock Journal of Medical Ethics. 2020;46(3):205--211.

\bibitem[{Guidotti et~al.(2018)Guidotti, Riccardo and Monreale, Anna and Ruggieri, Salvatore and Turini, Franco and Giannotti, Fosca and Pedreschi, Dino}]{guidotti2018survey}
Guidotti R, Monreale A, Ruggieri S, Turini F, Giannotti F, Pedreschi D.
\newblock A survey of methods for explaining black box models.
\newblock ACM Computing Surveys. 2018;51(5):1--42.

\bibitem[{Guo et~al.(2021)Guo, Lin Lawrence and Pfohl, Stephen R and Fries, Jason and Posada, Jose and Fleming, Scott Lanyon and Aftandilian, Catherine and Shah, Nigam and Sung, Lillian}]{guo2021systematic}
Guo LL, Pfohl SR, Fries J, Posada J, Fleming SL, Aftandilian C, et~al.
\newblock Systematic review of approaches to preserve machine learning performance in the presence of temporal dataset shift in clinical medicine.
\newblock Applied Clinical Informatics. 2021;12(4):808--815.

\bibitem[{Hatherley(2024)Hatherley, Joshua}]{hatherley2024data}
Hatherley J.
\newblock Data over dialogue: Why artificial intelligence is unlikely to humanise medicine.
\newblock PhD thesis, Monash University; 2024.

\bibitem[{Hatherley and Sparrow(2023)Hatherley, Joshua and Sparrow, Robert}]{hatherley2023diachronic}
Hatherley J, Sparrow R.
\newblock Diachronic and synchronic variation in the performance of adaptive machine learning systems: The ethical challenges.
\newblock Journal of the American Medical Informatics Association. 2023;30(2):361--366.

\bibitem[{Hatherley et~al.(2024)Hatherley, Joshua and Sparrow, Robert and Howard, Mark}]{hatherley2024virtues}
Hatherley J, Sparrow R, Howard M.
\newblock The virtues of interpretable medical AI.
\newblock Cambridge Quarterly of Healthcare Ethics. 2024;33(3):323--332.

\bibitem[{Hatherley(2020)Hatherley, Joshua James}]{hatherley2020limits}
Hatherley JJ.
\newblock Limits of trust in medical AI.
\newblock Journal of Medical Ethics. 2020;46(7):478--481.

\bibitem[{Hoernke et~al.(2021)Hoernke, Katarina and Djellouli, Nehla and Andrews, Lily and Lewis-Jackson, Sasha and Manby, Louisa and Martin, Sam and Vanderslott, Samantha and Vindrola-Padros, Cecilia}]{hoernke2021frontline}
Hoernke K, Djellouli N, Andrews L, Lewis-Jackson S, Manby L, Martin S, et~al.
\newblock Frontline healthcare workers’ experiences with personal protective equipment during the COVID-19 pandemic in the UK: A rapid qualitative appraisal.
\newblock BMJ Open. 2021;11(1):e046199.

\bibitem[{Jenkins et~al.(2021)Jenkins, David A and Martin, Glen P and Sperrin, Matthew and Riley, Richard D and Debray, Thomas PA and Collins, Gary S and Peek, Niels}]{jenkins2021continual}
Jenkins DA, Martin GP, Sperrin M, Riley RD, Debray TP, Collins GS, et~al.
\newblock Continual updating and monitoring of clinical prediction models: Time for dynamic prediction systems?
\newblock Diagnostic and Prognostic Research. 2021;5:1--7.

\bibitem[{Jesus et~al.(2021)Jesus, S{\'e}rgio and Bel{\'e}m, Catarina and Balayan, Vladimir and Bento, Jo{\~a}o and Saleiro, Pedro and Bizarro, Pedro and Gama, Jo{\~a}o}]{jesus2021can}
Jesus S, Bel{\'e}m C, Balayan V, Bento J, Saleiro P, Bizarro P, et~al.
\newblock How can I choose an explainer? An application-grounded evaluation of post-hoc explanations.
\newblock In: Proceedings of the 2021 ACM Conference on Fairness, Accountability, and Transparency; 2021. p. 805--815.

\bibitem[{Jonassen and Henning(1999)Jonassen, David H and Henning, Philip}]{jonassen1999mental}
Jonassen DH, Henning P.
\newblock Mental models: Knowledge in the head and knowledge in the world.
\newblock Educational Technology. 1999;39(3):37--42.

\bibitem[{Krishnan(2020)Krishnan, Maya}]{krishnan2020against}
Krishnan M.
\newblock Against interpretability: A critical examination of the interpretability problem in machine learning.
\newblock Philosophy \& Technology. 2020;33(3):487--502.

\bibitem[{Kroll(2018)Kroll, Joshua A}]{kroll2018fallacy}
Kroll JA.
\newblock The fallacy of inscrutability.
\newblock Philosophical Transactions of the Royal Society A: Mathematical, Physical and Engineering Sciences. 2018;376(2133):20180084.

\bibitem[{London(2019)London, Alex John}]{london2019artificial}
London AJ.
\newblock Artificial intelligence and black-box medical decisions: Accuracy versus explainability.
\newblock Hastings Center Report. 2019;49(1):15--21.

\bibitem[{McKernan and Davies(2024)Bethan McKernan and Harry Davies}]{mckernan2024machine}
McKernan B, Davies H.: "The machine did it coldly": Israel used AI to identify 37,000 Hamas targets; 2024.
\newblock \urlprefix\url{https://www.theguardian.com/world/2024/apr/03/israel-gaza-ai-database-hamas-airstrikes}.

\bibitem[{Miller(2021)Miller, Tim}]{miller2021contrastive}
Miller T.
\newblock Contrastive explanation: A structural-model approach.
\newblock The Knowledge Engineering Review. 2021;36:e14.

\bibitem[{Minh et~al.(2022)Minh, Dang and Wang, H Xiang and Li, Y Fen and Nguyen, Tan N}]{minh2022explainable}
Minh D, Wang HX, Li YF, Nguyen TN.
\newblock Explainable artificial intelligence: A comprehensive review.
\newblock Artificial Intelligence Review. 2022;55:1--66.

\bibitem[{Mitchell et~al.(2019)Mitchell, Margaret and Wu, Simone and Zaldivar, Andrew and Barnes, Parker and Vasserman, Lucy and Hutchinson, Ben and Spitzer, Elena and Raji, Inioluwa Deborah and Gebru, Timnit}]{mitchell2019model}
Mitchell M, Wu S, Zaldivar A, Barnes P, Vasserman L, Hutchinson B, et~al.
\newblock Model cards for model reporting.
\newblock In: FAT* '19: Proceedings of the Conference on Fairness, Accountability, and Transparency; 2019. p. 220--229.

\bibitem[{Molnar(2020)Molnar, Christoph}]{molnar2020interpretable}
Molnar C.
\newblock Interpretable machine learning.
\newblock Leanpub; 2020.

\bibitem[{Mosier et~al.(2017)Mosier, Kathleen L and Skitka, Linda J and Heers, Susan and Burdick, Mark}]{mosier2017automation}
Mosier KL, Skitka LJ, Heers S, Burdick M.
\newblock Automation bias: Decision making and performance in high-tech cockpits.
\newblock In: Decision Making in Aviation. Routledge; 2017. p. 271--288.

\bibitem[{Muralidharan et~al.(2024)Muralidharan, Anantharaman and Savulescu, Julian and Schaefer, G Owen}]{muralidharan2024ai}
Muralidharan A, Savulescu J, Schaefer GO.
\newblock AI and the need for justification (to the patient).
\newblock Ethics and Information Technology. 2024;26(1):16.

\bibitem[{Nguyen et~al.(2015)Nguyen, Anh and Yosinski, Jason and Clune, Jeff}]{nguyen2015deep}
Nguyen A, Yosinski J, Clune J.
\newblock Deep neural networks are easily fooled: High confidence predictions for unrecognizable images.
\newblock In: Proceedings of the IEEE Conference on Computer Vision and Pattern Recognition; 2015. p. 427--436.

\bibitem[{Poursabzi-Sangdeh et~al.(2021)Poursabzi-Sangdeh, Forough and Goldstein, Daniel G and Hofman, Jake M and Wortman Vaughan, Jennifer Wortman and Wallach, Hanna}]{poursabzi2021manipulating}
Poursabzi-Sangdeh F, Goldstein DG, Hofman JM, Wortman~Vaughan JW, Wallach H.
\newblock Manipulating and measuring model interpretability.
\newblock In: Proceedings of the 2021 CHI Conference on Human Factors in Computing Systems; 2021. p. 1--52.

\bibitem[{Pruski(2023)Pruski, Michal}]{pruski2023ethics}
Pruski M.
\newblock Ethics framework for predictive clinical AI model updating.
\newblock Ethics and Information Technology. 2023;25(3):48.

\bibitem[{Qui{\~n}onero-Candela et~al.(2022)Qui{\~n}onero-Candela, Joaquin and Sugiyama, Masashi and Schwaighofer, Anton and Lawrence, Neil D}]{quinonero2022dataset}
Qui{\~n}onero-Candela J, Sugiyama M, Schwaighofer A, Lawrence ND.
\newblock Dataset shift in machine learning.
\newblock MIT Press; 2022.

\bibitem[{Rudin(2019)Rudin, Cynthia}]{rudin2019stop}
Rudin C.
\newblock Stop explaining black box machine learning models for high stakes decisions and use interpretable models instead.
\newblock Nature Machine Intelligence. 2019;1(5):206--215.

\bibitem[{Rudin and Radin(2019)Rudin, Cynthia and Radin, Joanna}]{rudin2019we}
Rudin C, Radin J.
\newblock Why are we using black box models in AI when we don’t need to? A lesson from an explainable AI competition.
\newblock Harvard Data Science Review. 2019;1(2):1--9.

\bibitem[{Shen et~al.(2024)Shen, Judy Hanwen and Raji, Inioluwa Deborah and Chen, Irene Y}]{shen2024data}
Shen JH, Raji ID, Chen IY.
\newblock The data addition dilemma.
\newblock Proceedings of Machine Learning Research. 2024;252:1--43.

\bibitem[{Shortliffe and Sep{\'u}lveda(2018)Shortliffe, Edward H and Sep{\'u}lveda, Martin J}]{shortliffe2018clinical}
Shortliffe EH, Sep{\'u}lveda MJ.
\newblock Clinical decision support in the era of artificial intelligence.
\newblock JAMA. 2018;320(21):2199--2200.

\bibitem[{Singh et~al.(2022)Singh, Vinay and Chen, Shiuann-Shuoh and Singhania, Minal and Nanavati, Brijesh and Gupta, Agam and others}]{singh2022reinforcement}
Singh V, Chen SS, Singhania M, Nanavati B, Gupta A, et~al.
\newblock How are reinforcement learning and deep learning algorithms used for big data based decision making in financial industries -- A review and research agenda.
\newblock International Journal of Information Management Data Insights. 2022;2(2):100094.

\bibitem[{S{\o}gaard(2023)S{\o}gaard, Anders}]{sogaard2023opacity}
S{\o}gaard A.
\newblock On the opacity of deep neural networks.
\newblock Canadian Journal of Philosophy. 2023;53(3):224--239.

\bibitem[{Sparrow and Hatherley(2020)Sparrow, Robert and Hatherley, Joshua}]{sparrow2020high}
Sparrow R, Hatherley J.
\newblock High hopes for {“Deep Medicine”}? AI, economics, and the future of care.
\newblock Hastings Center Report. 2020;50(1):14--17.

\bibitem[{Sparrow et~al.(2024)Sparrow, Robert and Hatherley, Joshua and Oakley, Justin and Bain, Chris}]{sparrow2024should}
Sparrow R, Hatherley J, Oakley J, Bain C.
\newblock Should the use of adaptive machine learning systems in medicine be classified as research?
\newblock The American Journal of Bioethics. 2024;24(10):58--69.

\bibitem[{Sparrow and Hatherley(2019)Sparrow, Robert and Hatherley, Joshua James}]{sparrow2019promise}
Sparrow R, Hatherley JJ.
\newblock The promise and perils of AI in medicine.
\newblock International Journal of Chinese \& Comparative Philosophy of Medicine. 2019;17(2):79--109.

\bibitem[{Wachter et~al.(2017)Wachter, Sandra and Mittelstadt, Brent and Russell, Chris}]{wachter2017counterfactual}
Wachter S, Mittelstadt B, Russell C.
\newblock Counterfactual explanations without opening the black box: Automated decisions and the GDPR.
\newblock Harvard Journal of Law \& Technology. 2017;31(2):841--887.

\bibitem[{Wadden(2022)Wadden, Jordan Joseph}]{wadden2022defining}
Wadden JJ.
\newblock Defining the undefinable: The black box problem in healthcare artificial intelligence.
\newblock Journal of Medical Ethics. 2022;48(10):764--768.

\bibitem[{Zednik(2021)Zednik, Carlos}]{zednik2021solving}
Zednik C.
\newblock Solving the black box problem: A normative framework for explainable artificial intelligence.
\newblock Philosophy \& Technology. 2021;34(2):265--288.

\end{thebibliography}

\end{document}